\documentclass[journal]{vgtc}            % preprint (journal style)
\newcommand{\systemName}{{StuGPTViz}}

\title{{\systemName}: A Visual Analytics Approach to Understand Student-ChatGPT Interactions}
\author{
 \authororcid{Zixin Chen}{0000-0001-8507-4399},
 \authororcid{Jiachen Wang}{0000-0001-9630-9958},
 \authororcid{Meng Xia}{0000-0002-2676-9032},
 \authororcid{Kento Shigyo}{0000-0002-5095-7500},
 \authororcid{Dingdong Liu}{0000-0003-0985-0979},
 \authororcid{Rong Zhang}{0000-0002-4669-4816}, and 
 \authororcid{Huamin Qu}{0000-0002-3344-9694}
 }

\authorfooter{
  \item
  	Z. Chen, K. Shigyo, D. Liu, R. Zhang, J. Wang, H. Qu are with the Hong Kong University of Science and Technology, Hong Kong, China. E-mail: \{zchendf, kshigyo, dliuak, rzhangab\}@connect.ust.hk, csejiachenw@ust.hk and huamin@cse.ust.hk
  \item
  	M. Xia is the corresponding author, affiliated with Texas A\&M University, College Station, Texas, the United States. Email: mengxia@tamu.edu
}

\abstract{%
The integration of Large Language Models (LLMs), especially ChatGPT, into education is poised to revolutionize students' learning experiences by introducing innovative conversational learning methodologies. To empower students to fully leverage the capabilities of ChatGPT in educational scenarios, understanding students' interaction patterns with ChatGPT is crucial for instructors. However, this endeavor is challenging due to the absence of datasets focused on student-ChatGPT conversations and the complexities in identifying and analyzing the evolutional interaction patterns within conversations. To address these challenges, we collected conversational data from 48 students interacting with ChatGPT in a master's level data visualization course over one semester. We then developed a coding scheme, grounded in the literature on cognitive levels and thematic analysis, to categorize students' interaction patterns with ChatGPT. Furthermore, we present a visual analytics system, {\systemName}, that tracks and compares temporal patterns in student prompts and the quality of ChatGPT's responses at multiple scales, revealing significant pedagogical insights for instructors. We validated the system's effectiveness through expert interviews with six data visualization instructors and three case studies. The results confirmed {\systemName}'s capacity to enhance educators' insights into the pedagogical value of ChatGPT. We also discussed the potential research opportunities of applying visual analytics in education and developing AI-driven personalized learning solutions.

}


\teaser{
  \centering
  \includegraphics[width=\linewidth, alt={System interface of {\systemName}}]{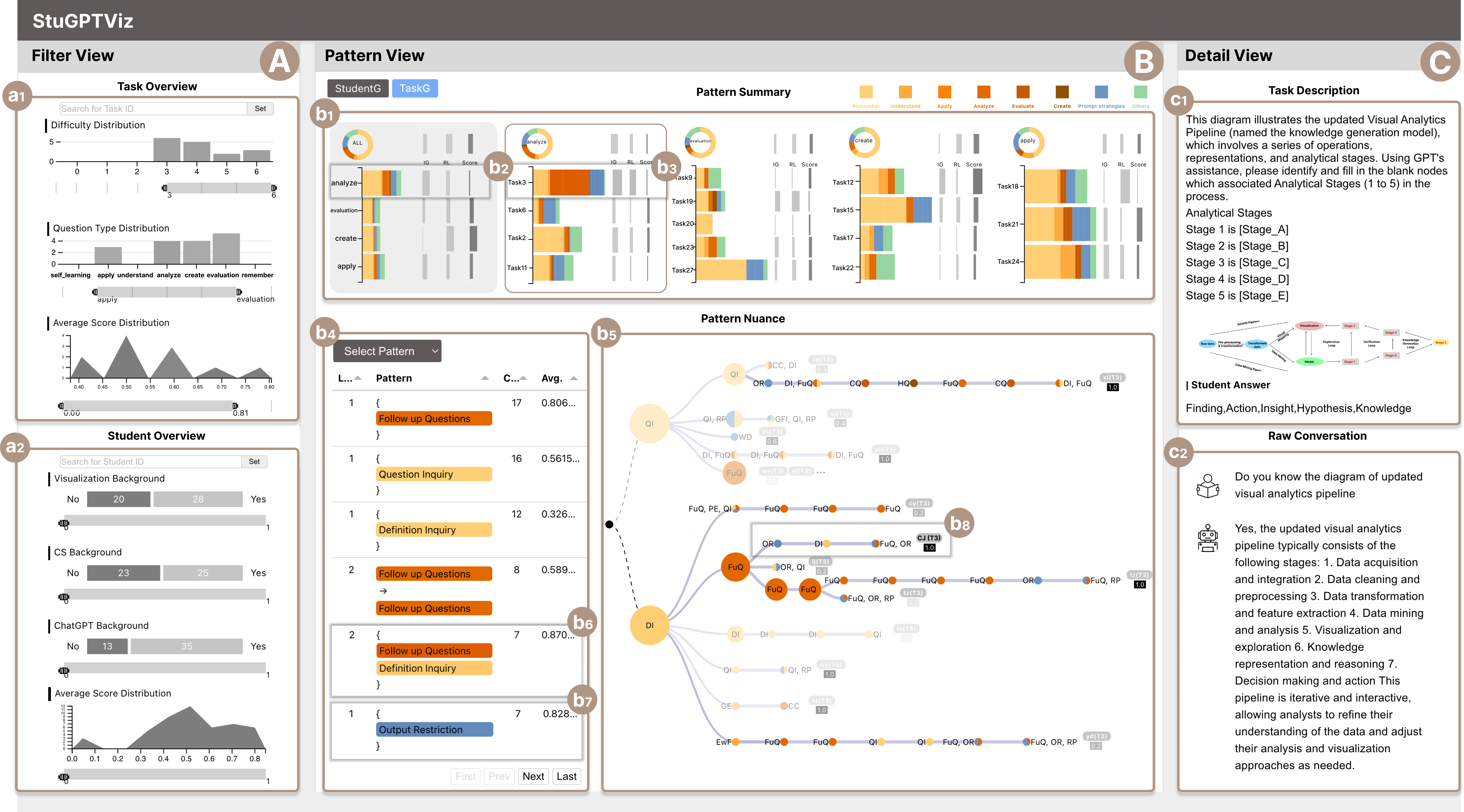}
  \caption{%
   System interface of {\systemName}. The \textbf{Filter View} (A) offers an overview and enables filtering of tasks and students through the Task Overview (a1) and Student Overview (a2). The \textbf{Pattern View} (B) displays the macro-level summary of conversation characteristics in the Pattern Summary (b1), and micro-level interaction patterns along with their evolution in the Pattern Nuance, represented by the Pattern Mining Table (b4) and the Interaction Tree (b5). The \textbf{Detail View} (C) presents task descriptions (c1), and the raw conversation data between students and ChatGPT (c2).
  }
  \label{fig:Teaser}
}

\graphicspath{{figs/}{figures/}{pictures/}{images/}{./}} % where to search for the images

\usepackage{tabu}                      % only used for the table example
\usepackage{booktabs}                  % only used for the table example
\usepackage{lipsum}                    % used to generate placeholder text
\usepackage{mwe}                       % used to generate placeholder figures
\usepackage{tabularx}

\usepackage{amsmath}
\usepackage{mathptmx}                  % use matching math font

\begin{document}

%%%%%%%%%%%%%%%%%%%%%%%%%%%%%%%%%%%%%%%%%%%%%%%%%%%%%%%%%%%%%%%%
%%%%%%%%%%%%%%%%%%%%%% START OF THE PAPER %%%%%%%%%%%%%%%%%%%%%%
%%%%%%%%%%%%%%%%%%%%%%%%%%%%%%%%%%%%%%%%%%%%%%%%%%%%%%%%%%%%%%%%

\firstsection{Introduction}

\maketitle

% The logic of intro: 

% New version:

% The emergence of artificial intelligence (AI) has catalyzed profound changes across numerous sectors, notably in education \cite{bonfield2020transformation}. 
Groundbreaking developments in generative AI, particularly through Large Language Models (LLMs) applications such as ChatGPT, have introduced unprecedented opportunities in educational methodologies \cite{bonfield2020transformation, grassini2023shaping}. These tools not only expedite students' information searches but also assist instructors in refining classroom activities and delivering personalized guidance\cite{lan2024teachers, bonner2023large, owan2023exploring}. However, as the integration of LLMs into educational scenarios is still nascent, it is imperative for instructors to carefully plan and assess how students utilize LLMs in their learning activities \cite{lodge2023mapping} to harness the full potential of LLMs and enhance the student learning experience \cite{gan2023large, tajik2024comprehensive}. A fundamental step in this process is to gain a comprehensive understanding of student interactions with LLMs, thereby acquiring key pedagogical insights such as students' cognitive levels, learning attitudes, and mastery of knowledge \cite{jeon2023large,kumar2023impact}.

% This insight is the cornerstone for leveraging AI tools like LLMs, ensuring they significantly enhance students' learning experiences and unlock broad-ranging benefits \placeholderRef{Requirement of understand how students use GPT}. 

% \revision{However, efforts to arm educators with this knowledge are still in the preliminary stages. Most research focuses on ethical concerns of using ChatGPT in educational environments or ways to integrate GPT into teaching methodologies, such as employing it as a teaching assistant. Detailed investigations into how students employ GPT in educational activities are markedly scant and encounter notable obstacles. First, there's a significant shortage of datasets on student-GPT conversations that are specifically aimed at educational contexts. Most existing datasets feature broad conversations with GPT across various topics, rather than educational dialogues.
% Second, analyzing extensive conversation datasets is challenging, distinct from traditional text analysis examining topics or semantics. 
% Student strategies should be identified and categorized from their dialogues with GPT to provide valuable insights for educators. The pattern of student strategies is an unknown area that little existing literature has investigated. Effective methods should be developed to summarize and highlight these valuable strategies.} 

Nevertheless, efforts to provide instructors with these pedagogical insights are still in the initial stages. Existing research on LLMs in education primarily focuses on educational ethics and potential application scenarios, such as using LLMs as automated tools for evaluating student essays \cite{rahman2023chatgpt,mhlanga2023open,imran2023analyzing,baidoo2023education,huallpa2023exploring}. To our knowledge, in-depth studies on students' interactions with LLMs for learning tasks are scarce, facing two main challenges. First, there are no publicly available datasets dedicated to capturing students' conversations with LLMs. The existing datasets are primarily composed of everyday conversations with general users. While some include conversations about learning tasks like programming or solving math problems, they are few and lack guaranteed quality because the learning scenarios involving LLM use are not carefully crafted and evaluated by instructors. Moreover, these conversations typically follow a ``one question, one answer'' format, as their primary purpose is to assess LLMs' problem-solving capabilities. These deficiencies highlight the urgent need for data collection of learning-centered conversations with well-structured tasks \cite{chen2023theoremqa,zheng2023lmsys,codealpaca}.

% A primary difficulty lies in accurately interpreting students' cognitive levels and thought processes based on their inquiries. \revision{This is a crucial question frequently raised by instructors in LLM-involved education scenarios, as they are eager to understand how much higher-order thinking (e.g., independent thinking) students are doing when using these advanced AI tools \cite{darvishi2024impact,vargas2023challenges}. }

Second, understanding how students interact with LLMs for pedagogical insights through conversation data presents significant challenges. One major difficulty is that instructors are eager to understand the extent of higher-order thinking (e.g., independent thinking) students engage in when using these advanced AI tools \cite{darvishi2024impact,vargas2023challenges}. To comprehend this higher-order thinking, it is essential to measure students' cognitive levels \cite{ball2005modeling}. However, accurately interpreting these cognitive levels based on students' inquiries to LLMs has never been explored. Additionally, assessing students' proficiency in utilizing LLMs poses another challenge, which involves evaluating the various LLMs' responses and observing how students adjust their prompts in response.
Moreover, tracking the progression of these interactions introduces an added layer of complexity \cite{faruk2023university, tajik2023comprehensive}. 
Visual analysis is a potential way. However, current research has often overlooked these challenges. While there are many visualizations works on conversation analysis studies focusing on topic progression or sentiment analysis \cite{hao2011visual, kwon2015visohc,6634160}, such studies do not adequately capture and visualize the cognitive levels reflected in the evolving interactions, falling short of meeting instructors' needs. 

To address these challenges, we selected ChatGPT, notable for being one of the most prevalent LLM applications \cite{10.1145/3649506}, to gather conversation data. In addition, we see a potential to introduce LLMs for visualization education to address diverse student backgrounds and manage varied learning activities such as concept comprehension, visualization literacy, and design\cite{lo2019learning}. In collaboration with experienced course instructors, We devised and integrated an in-class exercise module into a graduate-level data visualization course at the local university, allowing students to interact with ChatGPT freely. Our approach yielded a significant collection of high-quality student-ChatGPT conversation data from well-crafted learning tasks. Through comprehensive thematic analysis and an extensive literature review \cite{braun2012thematic,white2023prompt,shankar2024spade,white2023chatgpt}, we developed a coding scheme that categorizes the diverse cognitive levels \cite{krathwohl2002revision} and several metrics to evaluate the quality of ChatGPT responses \cite{es2023ragas,lee2006information}. To capture evolving strategies students use when interacting with ChatGPT, we analyzed various sequences and sets of the codes, defining these as ``interaction patterns'' which became the focal point of our analysis. Building on this foundation and design requirements derived from course instructors and experts, we introduce a pioneering visual analytics system for instructors to explore intricate interaction patterns and derive actionable pedagogical insights from student-ChatGPT conversation data. In particular, a customized tree visualization is designed to present the evolution and compare the characteristics of students' interaction patterns. To summarize, our key contributions are as follows:

\begin{itemize}[leftmargin=*,itemsep=0pt,topsep=0pt]
\item We introduced ChatGPT to a real data visualization course, collected student-ChatGPT conversation data and developed a coding scheme for an in-depth analysis of interaction patterns.
\item We designed a visual analytics system, {\systemName}, to help instructors discover insights into students' cognitive levels and proficiency when using ChatGPT.
\item Through three case studies and expert interviews, we demonstrated the effectiveness of our coding scheme and system in enhancing educational activities such as problem-solving guidance, personalized feedback, and exercise design.
\end{itemize}

% \begin{itemize}
% \item We collected high-quality student-GPT conversation data from an actual data visualization course and developed a coding scheme, enabling in-depth analysis of student-ChatGPT interaction patterns.
% \item We designed a visual analytics system that provides instructors with the capabilities to understand and analyze the nuances of students' evolving interaction patterns with ChatGPT and to clearly discern the underlying cognitive levels of students. This system aids instructors in conducting educational activities such as guiding students in using ChatGPT for problem-solving, providing personalized feedback for more customized education, and refining exercise designs to better meet educational goals.
% \item We conducted comprehensive evaluations through three case studies and expert interviews, which validates the effectiveness of our coding scheme and system.
% \end{itemize}

Overall, we present a design study that constitutes an initial yet crucial step toward analyzing student interaction patterns with ChatGPT, advancing the application of visual analytics in AI-driven education.

\section{Related Work}

% 1. ChatGPT for Edu. 2. VIS for Edu. 3. Event Sequence, Conversation and Tree Visualization

In this section, we discuss the relevant research, including LLMs in education and visualization education, visualization for AI-enhanced education, and visual analytics for conversational data.

\subsection{LLMs in Education and Visualization Education}
The integration of LLMs such as ChatGPT into educational settings has sparked a diverse range of discussions, with plenty of initial works centered on ethical considerations regarding their use in learning environments \cite{mhlanga2023open,huallpa2023exploring}. Increasingly, the academic community recognizes the transformative potential LLMs hold for education, advocating for their adoption to revolutionize learning and teaching methodologies \cite{bonfield2020transformation}. Despite potential resistance from some educators, students inevitably turn to LLMs for assistance with coursework \cite{grassini2023shaping}. Therefore, researchers and instructors have explored LLMs for various applications, including serving as teaching assistants for writing and coding, generating adaptive exercises~\cite{10.1145/3586182.3615785}, supporting personalized question-answering sessions~\cite{baidoo2023education}, and facilitating innovative learning modes like ``learn by teaching'', where LLM plays the role as ``learner" and students assume the role of the teacher to teach the AI~\cite{schmucker2023ruffle}.

However, there is a notable gap in understanding how students strategize their use of ChatGPT for educational purposes. Existing literature predominantly focuses on the capabilities and applications of LLMs without delving into student interaction strategies, leaving educators without the necessary insights to fully leverage these tools in enhancing learning experiences \cite{firat2023chat,abdelghani2023gpt}.

Simultaneously, the field of visualization education is gaining traction, not only within the visualization community but also more broadly \cite{bach2023challenges}. The challenges of teaching data visualization range from addressing diverse student backgrounds to managing varied learning activities such as concept comprehension, visualization literacy, and design evaluation—are substantial \cite{lo2019learning}. ChatGPT's potential to support these educational challenges opens avenues to investigate how students use ChatGPT across different visualization learning tasks, particularly relevant to our project's focus \cite{alafnan2023chatgpt}.
% Despite the growing interest in applying ChatGPT to teach data visualization, primarily centered around generating visualization-related questions, 
A comprehensive analysis of this research direction still remains unexplored \cite{cui2023adaptive}. This gap presents a unique opportunity for our work to contribute to the field by offering insights into student strategies in employing ChatGPT within visualization learning contexts, thereby advancing the understanding and application of LLMs in educational settings.

\subsection{Visualization for AI-Enhanced Education}
The integration of visualization in AI-enhanced education is dedicated to leveraging visual analytics for learning analysis, such as interpreting complex data and AI algorithms to improve educational outcomes \cite{duval2011attention}. While learning analysis harnesses data to refine and enhance learning processes, visualization techniques render these insights accessible and actionable for educators and students \cite{corrin2015loop,park2015development}. Despite notable advancements in each domain, the integration of visual analytics specifically tailored to learning analysis within AI-enhanced educational environments remains underexplored.

Existing research underscores the value of visual analytics in presenting student performance metrics, engagement levels, and learning behaviors, thus enriching our understanding of educational dynamics \cite{aljohani2019integrated,alhadad2018visualizing}. Within this context, the subfield of open learner models exemplifies the potential of visual explanations, akin to Explainable AI (XAI), in demystifying AI-generated outputs, offering learners and educators transparent and trustworthy insights \cite{bull2010open, epp2015uncertainty}. Additionally, other works have employed visual analytics to examine students' interaction data with intelligent agents, using students' log data such as hint requests to delve into their problem-solving processes \cite{xia2023involving}. Recently, with the advent of potent LLM-based conversational agents, the intricacy of interactions between students and AI has reached a new height. Goals once considered unrealistic, such as in-depth analysis of students' cognitive levels and thought processes, are now achievable \cite{javaid2023unlocking, wang2021towards, bai2023chatgpt}. To our knowledge, the dedicated exploration of visual analytics to analyze and elucidate student interactions with advanced AI tools, such as ChatGPT, is just beginning. 
% This recognized gap signifies an urgent need for research aimed at enhancing LLM-driven educational practices.
In response to this gap, our work proposes a novel visual analytics system based on the students-ChatGPT conversation data we collected. The system is designed to identify and unravel the intricate nuances of student-ChatGPT interactions. It equips educators with profound insights into how LLMs can be utilized to customize and elevate students' learning experiences.

\subsection{Visual Analytics for Conversational Data}
The exploration of visual analytics for conversational text data within the visualization community has encompassed a wide range of applications, from sentiment analysis and topic modeling to mapping conversation flows and interactions within user groups \cite{hao2011visual, kwon2015visohc,fu2016visual}. For instance, T-Cal and IneqDetect \cite{fu2018t,10.1145/3300115.3309528} focus on analyzing group conversations and estimating members' sentiments to assess collaboration effectiveness. Meanwhile, efforts such as ThreadReconstructor, MultiConVis, and VisOHC \cite{el2018threadreconstructor, hoque2016multiconvis, angus2011conceptual} probe into the structure and core topics of online forum discussions. However, these initiatives mainly focus on summarizing the dynamics of multi-party conversations without delving into the intricacies of one-on-one dialogues.

Another significant gap in current methodologies is their constrained ability to uncover the depth of evolving cognitive levels in educational dialogues between students and AI tools like ChatGPT. Visualization tools for one-on-one medical conversations, like ConVIScope and Discursis \cite{li2021conviscope, angus2012visualising}, focus on charting patient-doctor dialogues but only reflect changes in sentiment and topic over time. They fall short in showcasing how one party (e.g., students) adjusts their responses to another party (e.g., LLMs' replies). Similarly, research aimed at analyzing educational dialogues often emphasizes engagement and comprehension, lacking a detailed visual analysis of the depth of thinking, learning strategies, or intentions revealed through these interactions \cite{littleton2010educational,zuniga2002intergroup, bach2023challenges}.

These shortcomings highlight the necessity for a novel visual analytics framework designed to tackle the specific challenges posed by educational dialogues with LLMs \cite{fuchs2023exploring,alafnan2023chatgpt}. Consequently, our work introduces multiple visualizations, such as the Interaction Tree, to meticulously analyze students' interaction patterns with ChatGPT.
\section{Background and Data Collection}
\label{sec:requirement}
This section outlines the background of our data visualization course, the collaborative efforts with our expert team, the in-class exercises, and the data collection considerations and procedures we adopted.

\subsection{Data Collection Background and Considerations}
At the invitation of our experts (two course instructors, E1 \& E2), we set out to incorporate ChatGPT into the curriculum of a postgraduate data visualization course for computer science majors in the first semester of 2024. This course, meeting once a week for three hours, attracted 55 registrants. Over the past four months, we have collaborated closely with E1, E2, and three teaching assistants (TA1-TA3) at our university. E1 is a professor with over 15 years of experience designing and teaching data visualization courses. E2 is a lecturer with three years of teaching experience and served as the primary instructor for this course. TA1, TA2, and TA3, are senior teaching assistants who have supported the professors in designing the in-class exercises, homework, and exams for the data visualization course for at least two semesters. All experts have used ChatGPT extensively over the last two years. In preparation, we conducted a three-hour meeting to outline how ChatGPT would be integrated into the course and how data would be collected. To maximize the benefits of ChatGPT for students, we decided to introduce an in-class exercise section. This would allow students to apply what they learned by interacting with ChatGPT and create an ideal setting for collecting diverse data on their learning interactions. During our discussion, two key considerations emerged:

\textbf{C1: Diverse Learning Tasks for Comprehensive Data Collection.} To capture the full spectrum of student interactions with ChatGPT, exercises should be designed across various course aspects using diverse learning tasks (e.g., visualization understanding \& design). These learning tasks should encompass different cognitive levels as outlined in Bloom's Taxonomy \cite{krathwohl2002revision} to ensure that the collected data fully reflects students' diverse engagement with ChatGPT. While these tasks are exemplified using a data visualization course, learning tasks for any course can be similarly designed to align with Bloom's Taxonomy \cite{crowe2008biology}.

% Additionally, the learning tasks should cover different cognitive levels as per Bloom's Taxonomy \cite{krathwohl2002revision} to ensures that the collected data reflects a wide range of learning activities, enabling a deep understanding of students' engagement with ChatGPT.}  
% With comprehensive conversation data, we can deeply understand student's learning activities and engagement with ChatGPT. 

% \item 
\textbf{C2: Natural Learning Scenarios for Unbiased Data Collection.} To accurately reflect genuine students' behaviors, it was vital to integrate ChatGPT into the learning process as an optional tool rather than a course mandate. This approach was intended to gather authentic interaction data, showcasing real student needs and preferences in using ChatGPT for learning.
% \end{enumerate}

% Guided by these considerations, we incorporated a 40-minute open-ChatGPT session towards the end of each class. This session included multiple tasks designed to motivate students to engage actively with ChatGPT. We also conducted a pilot experiment to simulate the entire in-class exercise section to test the procedure and the design of the exercise questions with around 30 computer science Ph.D. students and collected their conversation data to confirm the suitability of our data collection process settings. The data collection process has also been approved by the university’s ethics assessment (IRB approved). In the following sections, we briefly introduce our in-class exercise task designs and the data collection procedure.

% \subsubsection{Tasks and In-class Exercise Procedure}
% Together with the instructors E1 \& E2, we designed 27 exercise tasks which diverse seven categories (\textbf{C1}) as follows:

Informed by these considerations, we included a 40-minute open ChatGPT session at the end of each class (9 classes in total) to promote active student engagement with ChatGPT through various tasks. Additionally, we carried out a pilot study with about 30 HCI and data visualization Ph.D. students to evaluate the exercise section's design and procedures. Their conversation data helped us ascertain the appropriateness of our data collection approach. The data collection process was also approved by the university's ethics committee (IRB approval). The subsequent sections will provide an overview of our task designs for the in-class exercises and the data collection procedures.

\subsection{Tasks Summary and In-class Exercise Procedure}
Collaborating with instructors E1 and E2, we developed 27 exercise tasks spread across seven distinct types according to the levels of cognitive learning in the Revised Bloom's Taxonomy (\textbf{C1}, \cref{fig:tasks}):

\begin{figure}[!htbp]
    \centering
    \includegraphics[width=\linewidth]{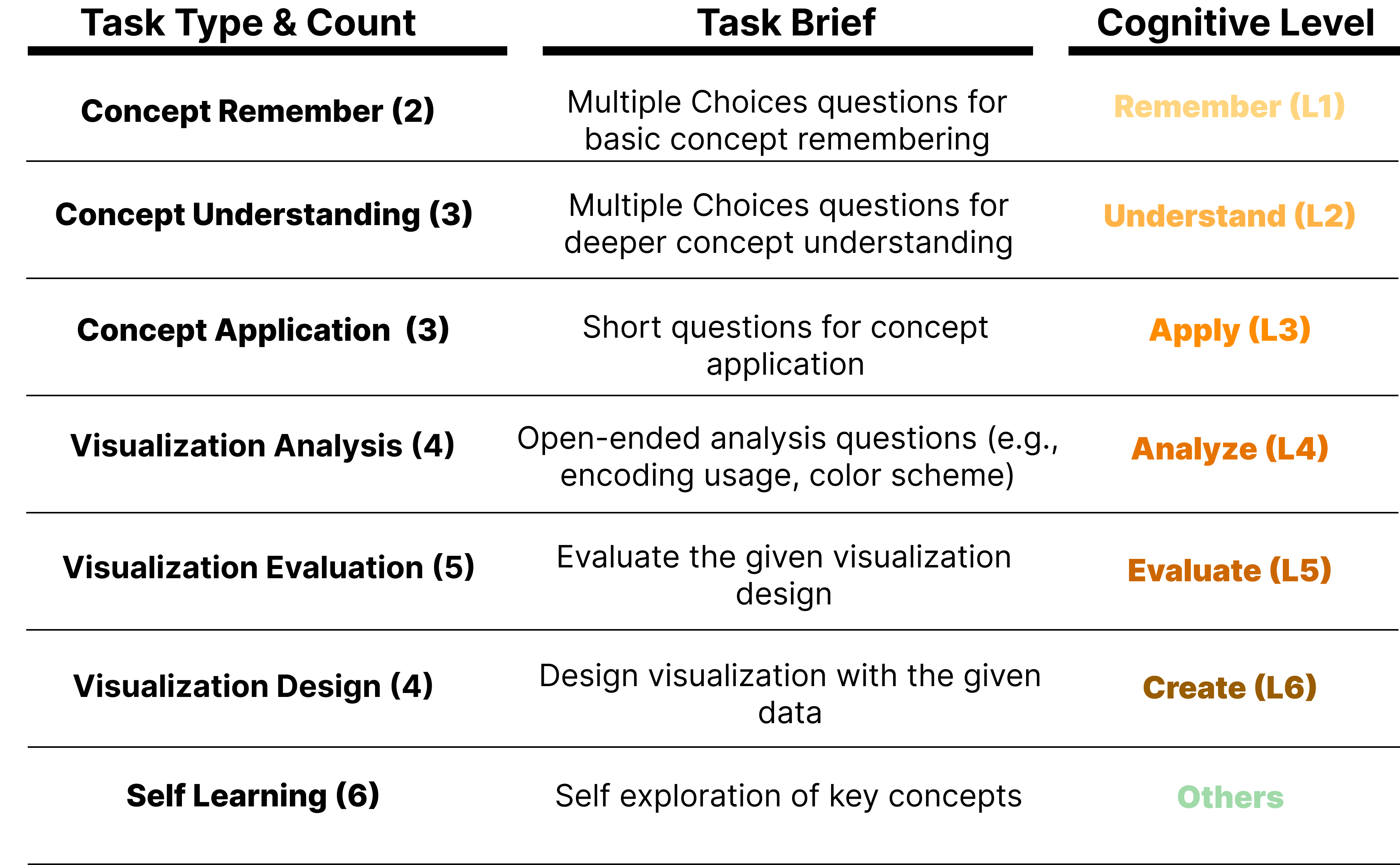}
    \caption{The summary of task type, count, cognitive level, and a brief description. Sample tasks are provided in the supplementary (A).}
    \label{fig:tasks}
\end{figure}

% \begin{table}[ht]
% \centering
% \caption{In-class Exercise Tasks Summary}
% \label{tab:exercise_tasks}
% \begin{tabularx}{\linewidth}{|>{\centering\arraybackslash}X|>{\centering\arraybackslash}X|}
% \hline
% \textbf{Task Type \& Count} & \textbf{Task Brief} \\ \hline
% Self Learning (6) & Self exploration of key concepts \\ \hline
% Concept Remember (2) & Multiple choices question for concept remembering \\ \hline
% Concept Understand (3) & Multiple choices question for concept understanding. \\ \hline
% Concept Apply (3) & Short questions for concept application. \\ \hline
% VIS Analyze (4) & Open-ended questions for visualizations analysis \\ \hline
% VIS Evaluate (5) & Evaluate the visualization design. \\ \hline
% VIS Design (Create) (4) & Design visualizations for the give data. \\ \hline
% % Add more rows as needed
% \end{tabularx}
% \end{table}

% To begin the data collection, we distributed a questionnaire to the 55 enrolled students to gauge their willingness to participate and to gather their background information. 48 students consented to participate, providing details on their undergraduate majors, data visualization expertise, programming skills, and prior experience with ChatGPT. In addition, we hosted a 40-minute introductory session on ChatGPT for beginners, which included instructions on how to export and upload the conversation files to the Canvas 

To begin data collection, we distributed a questionnaire to the 55 enrolled students to gauge their willingness to participate and gather background information. Forty-eight students consented, providing details on their undergraduate majors, data visualization expertise, programming skills, and prior ChatGPT experience. We also hosted a 40-minute introductory session on ChatGPT, including instructions on exporting and uploading conversation files to Canvas system\footnote{\url{https://www.instructure.com/canvas}}, the web-based learning management system used by our university.

Each in-class exercise session, conducted during the last 40 minutes of the lecture, included a 10-minute self-learning segment with ChatGPT, a 25-minute task completion segment, and a 5-minute conversation log upload phase. Students first spent 10 minutes asking ChatGPT questions to learn key lecture terms. This was followed by 25 minutes of task completion (\cref{fig:tasks}) with ChatGPT's assistance and a five-minute interval for uploading conversations to Canvas. Students were encouraged to interact naturally with ChatGPT (\textbf{C2}), and those confident with the course material were free skip the self-exploration.

% Each in-class exercise session, conducted during the last 40 minutes of the lecture, consisted of a self-learning segment with ChatGPT (10 mins), a task completion segment with ChatGPT (25 mins), and a conversation log upload phase (5 mins). Students initially were required to spend 10 minutes asking ChatGPT questions to learn the lecture's key terms. This was followed by a 25-minute segment where students completed given tasks (\cref{fig:tasks}) with ChatGPT's assistance and concluded with a five-minute interval for uploading their conversations to Canvas. We encouraged students to interact with ChatGPT in a way that felt natural to them. We suggested that those who were already confident with the course material could skip the self-exploration phase (\textbf{C2}) if they found it to be redundant.

\subsection{Dataset Brief}
The dataset we collected involved 48 students' conversation data with ChatGPT during the in-class exercise session of data visualization course over the entire spring 2024 semester. It consists of 744 unique conversations with 2507 turns after filtering out the empty conversations and those unrelated to the learning tasks. To the best of our knowledge, it is the only existing dataset specifically for student-ChatGPT conversations under strictly-defined learning activities. The collected data is in a structured format, including metadata such as student names (anonymized), task ID and task types, and conversation content. Each conversation is logged in sequential order, capturing both student prompts and ChatGPT responses. Sample data are provided in the supplementary materials. Additionally, we collected students demographic information, including their background in computer science, data visualization, and ChatGPT usage experience.

\section{Design Requirement and Data Processing}
This section presents the design requirements for the visual analytics system identified through discussions with our experts (E1 \& E2), the procedure for coding student prompts, and the methodology for processing ChatGPT responses.

\subsection{Visualization Design Requirements}
\label{design_requirements}
We summarized experts' requirements for analyzing student-ChatGPT conversation data as follows:
% \begin{enumerate}[label=\textbf{R{\arabic*}}, nolistsep]

% \item 
\textbf{R1: Overview of students and tasks data.} Experts highlighted the necessity of an overview of both students and the tasks, including the distribution of students' background information (e.g., knowledge in visualization) and tasks' characteristics (e.g., types). Instructors can then select specific students or tasks for deeper analysis of student-ChatGPT conversations.

% \item 
\textbf{R2: Summarizing macro-level conversation characteristics.} Before detailed analysis, experts require a comprehensive summary of the student-ChatGPT conversation, especially the cognitive levels of students' prompts and the qualities of ChatGPT responses based on selected tasks and students. This summary should span multiple user-selected viewpoints, covering broad categories such as various student groups and types of tasks, since instructors are always interested in the differences in behavior and performance among different student groups, such as those with and without a CS background. Additionally, instructors often have limited time and want to quickly see the differences among groups to prioritize their focus. 

\textbf{R3: Identifying micro-level interaction patterns.} Experts require a structured method to detect and summarize students' micro-level interaction patterns (i.e., recurring methods and strategies students employ when interacting with ChatGPT), along with essential metrics such as learning outcomes and pattern frequency. Emphasizing this information is crucial for instructors to identify which patterns are more effective for learning and merit deep exploration. They can further develop actionable insights on how students engage with ChatGPT throughout their learning journey.
% , thus optimizing the educational experience.

% \item 
\textbf{R4: Tracing interaction pattern evolution.} Educators necessitate a method to trace the development of students' interaction patterns, emphasizing both the shared and unique sequences within the context of task-solving. This requirement involves visualizing how students' interaction patterns evolve in response to tasks, reflecting variations in cognitive engagement and problem-solving approaches. Such visualization should facilitate a deeper understanding of varied student approaches to learning tasks.

% \item 
\textbf{R5: Evaluating interaction pattern performance.} Experts wanted to evaluate interaction patterns of interest effectively. This involves identifying students who utilize the pattern and assessing relevant metrics such as their learning outcomes and ChatGPT's response quality. Such comprehensive analysis helps gauge the effectiveness of different interaction patterns, enabling instructors to provide targeted feedback and identify exemplary patterns as recommended learning strategies.

% \item 
\textbf{R6: Examining detailed raw data.} 
Experts expressed a desire to access the raw data, which includes original in-class activities, students' answers or responses, and students' conversation logs with ChatGPT. These details can be used to justify their analysis results and provide straightforward examples for students to master effective interaction with ChatGPT.

\subsection{Students' Prompts Coding}
\label{prompt_coding}
The open coding process for students' prompts is based on thematic analysis methodology \cite{vaismoradi2013content} and enriched by a literature review to identify prompt patterns \cite{white2023prompt,white2023chatgpt,shankar2024spade}. Following expert requirements (\textbf{R2}), we categorized codes into two types: learning-related codes reflecting students' cognitive levels regarding course materials, and ChatGPT-related codes denoting students' comprehension and proficiency in using ChatGPT as a supplementary tool.
Firstly, we reviewed literature analyzing general user prompt patterns and intents \cite{white2023prompt,white2023chatgpt,shankar2024spade}, identifying 16 general prompt patterns as the initial ChatGPT-related codes. While these codes could represent students' proficiency with ChatGPT, they did not capture their cognitive levels. Therefore, we invited three visualization researchers, each with over four years of experience in data visualization education and thematic analysis, to independently conduct open coding of students' prompts. They refined the ChatGPT-related codes and developed learning-related codes that encapsulate the intent and thought processes behind each student's prompt. Our researchers engaged in repeated cross-checking and refinement until a consensus was reached, ultimately establishing 27 codes, including 15 learning-related and 12 ChatGPT-related codes.

\begin{figure}[!htbp]
    \centering
    \includegraphics[width=\linewidth]{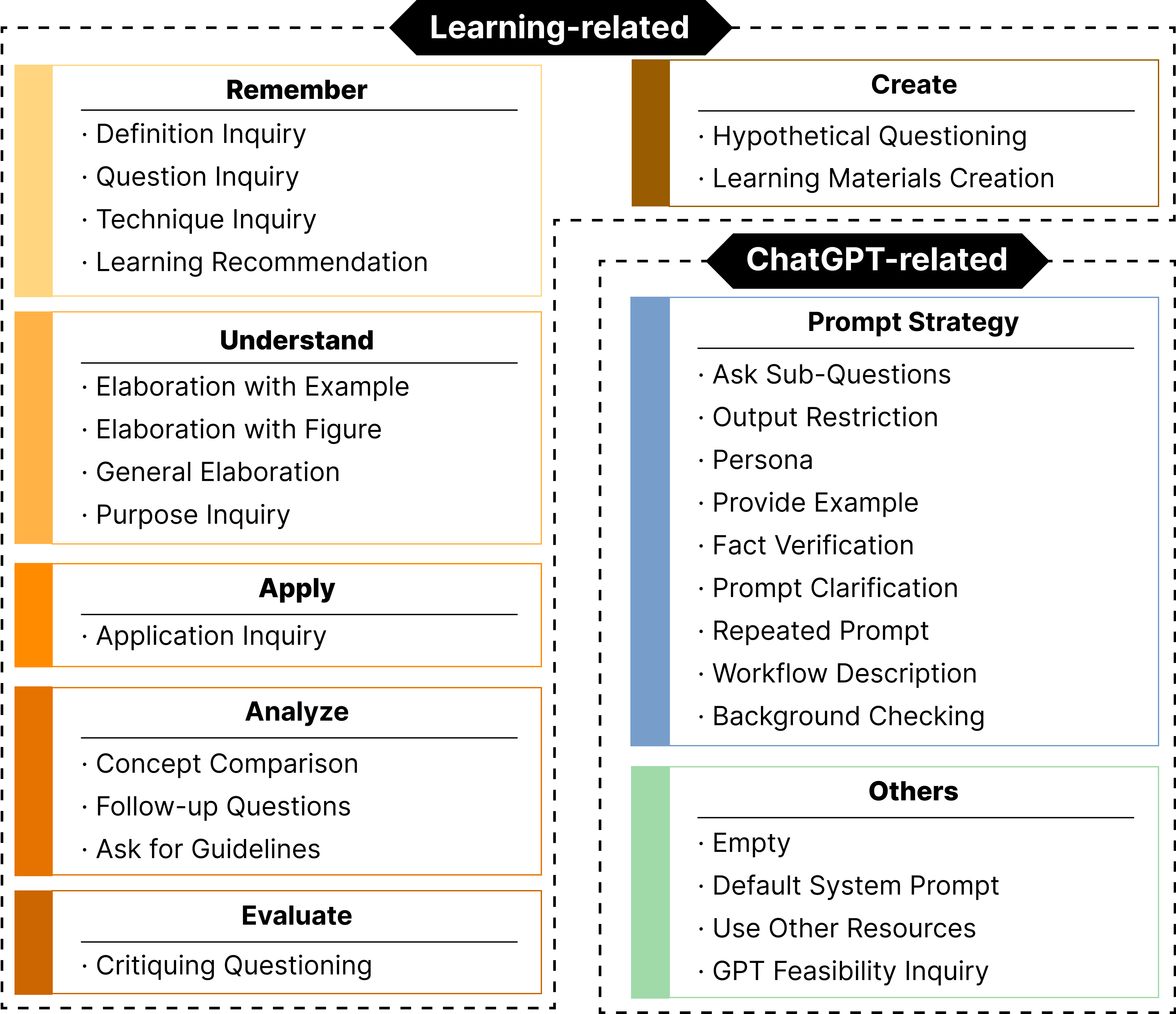}
    \caption{The code schema with revised bloom taxonomy \cite{krathwohl2002revision} classification. }
    \label{fig:code schema}
\end{figure}

\begin{figure*}[!htbp]
    \centering
    \includegraphics[width=\textwidth]{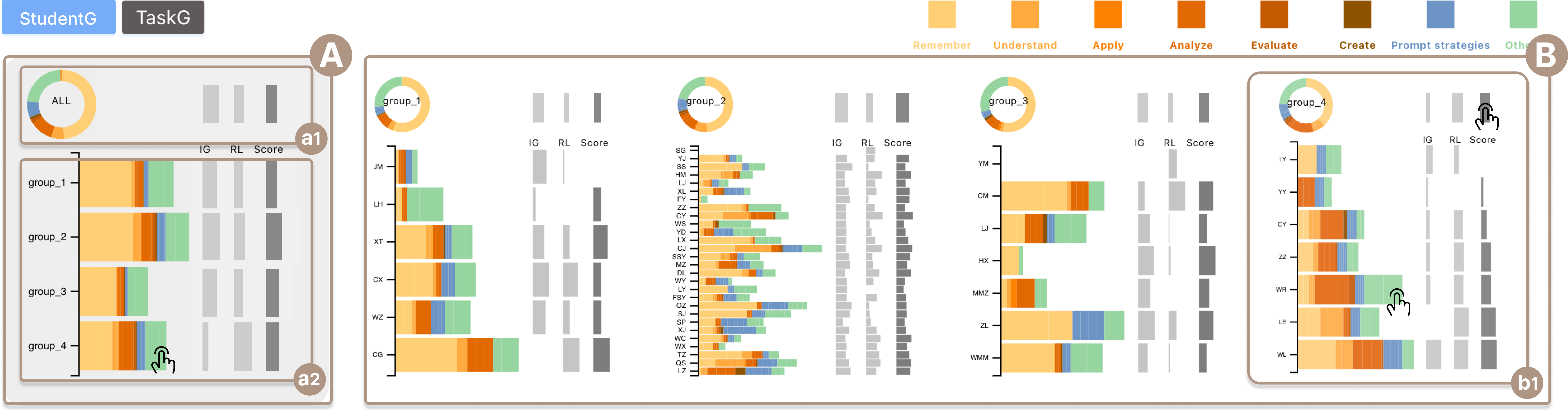}
    \caption{The \textbf{Pattern Summary} section of the \textit{Pattern View} summarizes both between-group and within-group interaction patterns based on the students and tasks selected by the user. Users can click on each grey bar to sort the students according to the selected metric.}
    \label{fig:overallCards}
\end{figure*}

After finalizing the code space, we consulted with our experts (TA1-TA3) to further refine our codes. At TA3's suggestion, we used the revised Bloom's taxonomy to categorize the 15 learning-related codes, enhancing our understanding of students' learning intentions and mapping their cognitive processes \cite{krathwohl2002revision}. This taxonomy describes six stages of cognitive learning: remember, understand, apply, analyze, evaluate, and create, each representing an advancement in cognitive level. The authors independently categorized the learning-related codes, then discussed and refined any inconsistencies until agreement was reached. The final code schema is summarized in \cref{fig:code schema}. To ensure coding quality and consistency, the authors coded 30 conversations from the pilot experiment with the finalized labels and calculated the Inter-Rater Reliability (IRR) scores, resulting in an IRR of 0.84, validating the coding process's reliability. Finally, the authors coded all students' prompts. For prompts embodying multiple learning intents and strategies, we applied multiple codes. Additionally, we used the ordered sequence of codes and unordered sets of codes from each student's conversation data to illustrate the ``interaction patterns'' identified via expert requirements (\textbf{R3}). For each coded conversation, we mined all possible ordered code sequences (e.g., $[\text{Definition Inquiry}, \text{Follow up Question},... ]$) and unordered code sets (e.g., $ \{\text{Definition Inquiry}, \text{Application Inquiry},... \}$) for further analysis.

\subsection{Processing ChatGPT's Responses}
\label{gpt_metrics}
To enhance our understanding of students' interactions with ChatGPT, we analyze and evaluate ChatGPT's response quality based on expert suggestions (TA1). Through our literature review, common metrics for evaluating ChatGPT's responses include response relevance, length, and correctness \cite{fergus2023evaluating,kocon2023chatgpt,johnson2023assessing}, all recognized as important by our experts. We employed the Ragas response relevance package \cite{es2023ragas}, a renowned framework for evaluating large language models, to assign a numerical score to the relevance of each ``user prompt - LLM response'' pair. Additionally, we measured response length and, in collaboration with experts E2 and TA2, assessed response correctness by categorizing them into ``basically correct'' (score $1$), ``partially correct'' (score $0.5$), and ``basically wrong'' (score $0$), respectively. Through further discussion, experts (E1 \& E2) expressed interest in evaluating the amount of accurate information a student can acquire from each turn of interaction with ChatGPT. Consequently, we combine the two metrics, response relevance score and response correctness, to develop the new metric ``information gain'' inspired by the literature \cite{liu2003kullback}. The metric's formula is shown below, which primarily leverages the KL-divergence principle \cite{van2014renyi}, calculates the amount of new information provided by the latest ChatGPT response compared to the existing set of responses:

\vspace{-2mm}
\[
\text{IG}(P, Q) = \sum_{i} P(i) \log \left( \frac{P(i)}{Q(i)} \right) \times R \times C,
\]

\vspace{-2mm}

In this formula, $IG$ represents the information gain of the incoming response $P$ under the cumulative response set $Q$. $P(i)$ is calculated as the frequency of word $i$ in the current response divided by the total number of words in that response. $Q(i)$, on the other hand, is the cumulative frequency of word $i$ up to the current response, divided by the total number of words in all responses up to that point. The variables $R$ and $C$ represent the numeric relevance score and correctness score, respectively. This computation quantifies the new information provided by ChatGPT's latest response compared to the existing knowledge base. 

\section{Visualization}
% Based on the derived design requirements in Section Three, we develop a visual analytics system, {\systemName} (\cref{fig:Teaser}), to help instructors analyze students' interaction with ChatGPT in our data visualization course. In this section, we will first provide an overview of the system. Then, we will provide a  detailed descriptions of each view's visualization design, corresponding interactions, and a discussion about potential alternative designs. 
Based on the design requirements identified in \cref{design_requirements} and the data collected in \cref{sec:requirement}, we developed a visual analytics system, {\systemName} (\cref{fig:Teaser}), aimed at enabling instructors to analyze student interactions with ChatGPT effectively. 
% This section is structured as follows: we begin with a system overview, followed by detailed descriptions of the visualization design for each view, together with the explanation of supported interactions and the discussion on potential alternative designs. \ToDo{add alternative sankey design for the tree}

\subsection{System Overview}

{\systemName} is intricately designed to facilitate a multi-level analysis of students' interactions with ChatGPT from the perspective of both tasks and students. It supports instructors in selecting specific tasks and students as focal points of analysis, thereby accommodating diverse analytical interests. Moreover, it enables a ``gradually deepening'' analysis process, allowing users to acquire both a broad overview and detailed insights into the interactions between students and ChatGPT. Specifically, {\systemName} is structured into three main components: 

% \ToDo{Whether a figure to show system workflow is needed? Mainly about frontend. If the backend data processing procedure is needed, should it be added to Chap.4?}

\textbf{Initial selection:} Instructors begin by selecting particular tasks or students of interest through the \textit{Task Overview} and \textit{Student Overview} in the \textbf{Filter View} (\cref{fig:Teaser}-a1, a2). This selection offers an overview of students and tasks data \textbf{(R1)}, which also triggers updates in the \textit{Pattern Summary} of the \textit{Pattern View} (\cref{fig:Teaser}-b1).

\textbf{Gradually deepening exploration:} Starting from the \textit{Pattern Summary} in the \textbf{Pattern View} (\cref{fig:Teaser}-b1), instructors can gain a summary of macro-level conversation characteristics of the selected students and tasks, covering both groups and individuals \textbf{(R2)}. To delve deeper, instructors can further analyze the micro-level student-ChatGPT interaction patterns via \textit{Pattern Nuance}. While the \textit{Pattern Mining Table} (\cref{fig:Teaser}-b4) enable instructors to identify the pattern summary together with significant metrics like learning outcome \textbf{(R3)}, the \textit{Interaction Tree} traced the pattern evolution of each student \textbf{(R4)}. Moreover, the interplay between these two visualizations enables instructors to explore and assess patterns of interest effectively \textbf{(R5)}.

\textbf{Detailed inspection:} Finally, the \textit{Task Description} and \textit{Raw Conversation} from the \textbf{Detailed View} (\cref{fig:Teaser}-c1, c2) provides access to the original tasks, students' responses, and their raw conversation logs with ChatGPT. This component allows instructors to examine task specifications, student prompts, and ChatGPT's replies, leveraging their expertise to interpret the data comprehensively \textbf{(R6)}.

% Following the aggregated overview, instructors can delve deeper into the analysis via \textit{Pattern View} (\cref{fig:Teaser}\textbf{b2}). 

% by exploring detailed interaction patterns in the lower section of the . This phase is designed for a more granular examination and comparison of different interaction dynamics and learning outcomes.

% By summarizing and detailed presenting the interaction patterns  task categories, student backgrounds, and interaction effectiveness, the system presents a comprehensive analysis to assist educators in making data-driven decisions to enhance learning outcomes. 

% It consists of three main components: initially, the user selects a task, student, or group of interest; this leads to an overview analysis at both the group and individual levels of student's interactions with ChatGPT. After acquiring a summarized analysis of the selected students or tasks, users can further conduct in-depth exploration, analysis, and comparison of detailed interaction patterns. Lastly, detailed examples are provided, including contextual data such as original task descriptions and students' answers, to offer users concrete examples for applying their insights effectively.

\subsection{Filter View}
The \textit{Filter View} (\cref{fig:Teaser}-A) serves as the gateway to analysis, presenting the distribution of various background metrics and enabling instructors to filter the students and tasks of interest (\textbf{R1}). The \textit{Task Overview} (\cref{fig:Teaser}-a1) displays information about learning tasks and features a search box for quickly finding specific tasks by ID. The distribution of task difficulties and types, determined by experts (E1 \& E2), is depicted through two bar charts, while an area chart illustrates the distribution of students' normalized average scores (x-axis) across tasks. Instructors can select tasks by clicking on the bars or adjusting sliders, with each metric chart dynamically updating in response to real-time user selections. Moreover, the \textit{Student Overview} (\cref{fig:Teaser}-a2) provides insights into students' backgrounds. A search box allows quick location of students by aliases, and an area chart at the bottom visualizes the distribution of students' average scores. Positioned between them, three-segmented bar charts represent the distribution of students' prior experience in data visualization and computer science and their familiarity with ChatGPT, derived from a background survey conducted during the first class. By freely filtering each metric, instructors can effortlessly isolate students and tasks of interest (\textbf{R1}), then proceed to the \textit{Pattern View} (\cref{fig:Teaser}-B). 

\subsection{Pattern View}
After identifying tasks and students of interest, instructors can delve into detailed analysis using the \textit{Pattern View} (\cref{fig:Teaser}-B). Comprising the \textit{Pattern Summary} and \textit{Pattern Nuance}, the \textit{Pattern View} enables a gradually deepening examination of interaction patterns, offering insights from macro-level summaries to micro-level details \textbf{(R2, R3)}. Throughout the \textit{Pattern View}, we consistently apply color scheme (\cref{fig:Teaser}, top-right corner) to represent the code categories defined in \cref{fig:code schema}. Specifically, we use a sequential color scheme, transitioning from light yellow to dark brown, to denote increasing cognitive levels (from ``remember'' to ``create'') as demonstrated in students' prompts. Concurrently, a distinct color scheme (blue for effective prompt strategies from literature and green for others) signifies students' proficiency in using ChatGPT.

\begin{figure}[!htbp]
    \centering
    \includegraphics[width=\linewidth]{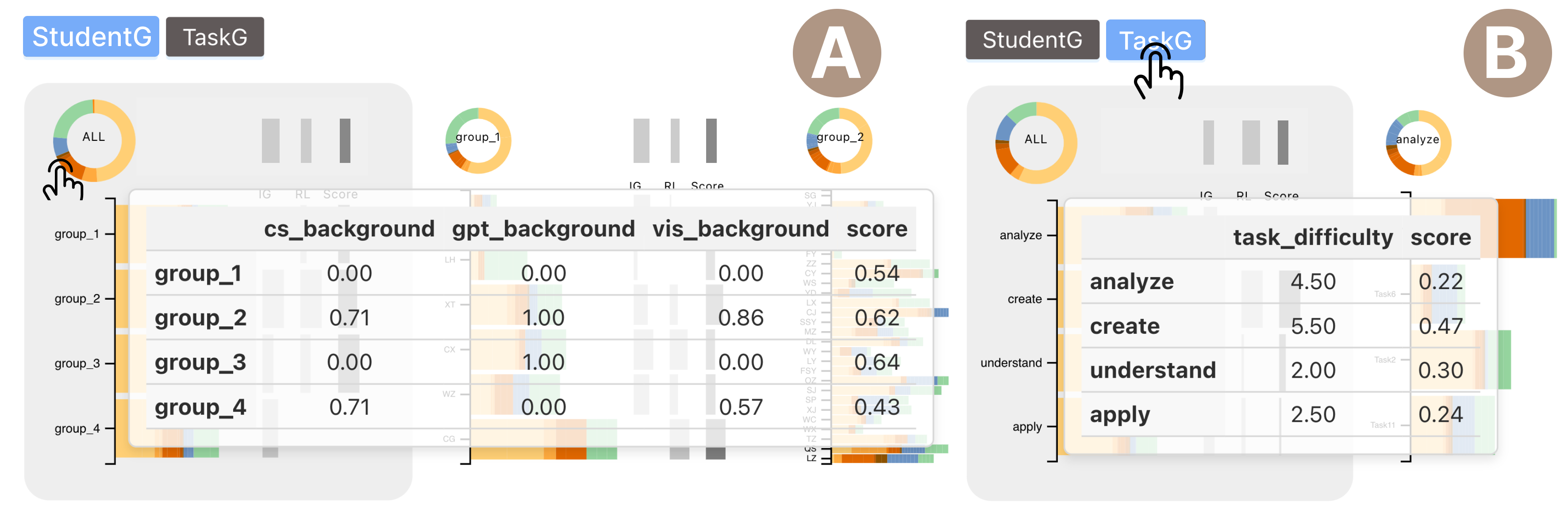}
    \caption{ (A) The between group-level background comparison. By default, students are grouped by their background. (B) Under the ``Task-Grouping'' mode, tasks are grouped by types.}
    \label{fig:group comparison & hover}
\end{figure}

\subsubsection{Pattern Summary}
\begin{figure}[!htbp]
    \centering
    \includegraphics[width=\linewidth]{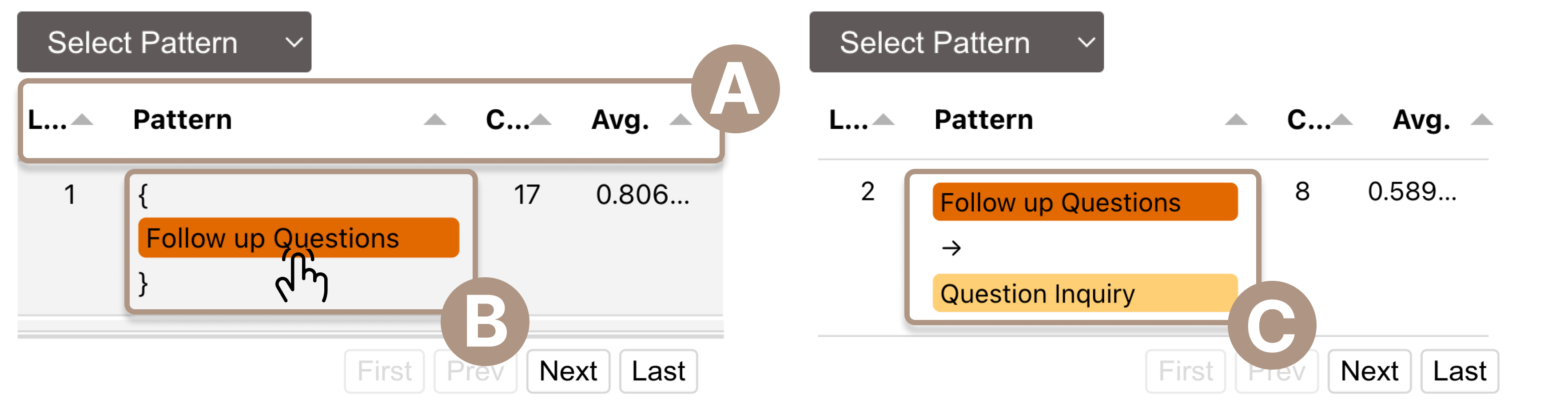}
    \caption{ The \textbf{Pattern Mining Table} within \textit{Pattern Nuance}. (A) The table headers include pattern length (``L''), interaction pattern (``Pattern''), pattern frequency (``C''), and average score (``Avg.''). (B) Example of the ``unordered code set'' type pattern. Users can click the pattern row to highlight the students utilizing this pattern in (\cref{fig:TreeDesign}-A). (C) Example of the ``ordered code list'' type pattern.}
    \label{fig:tableVis}
\end{figure}

\begin{figure*}
    \centering
    \includegraphics[width=\textwidth]{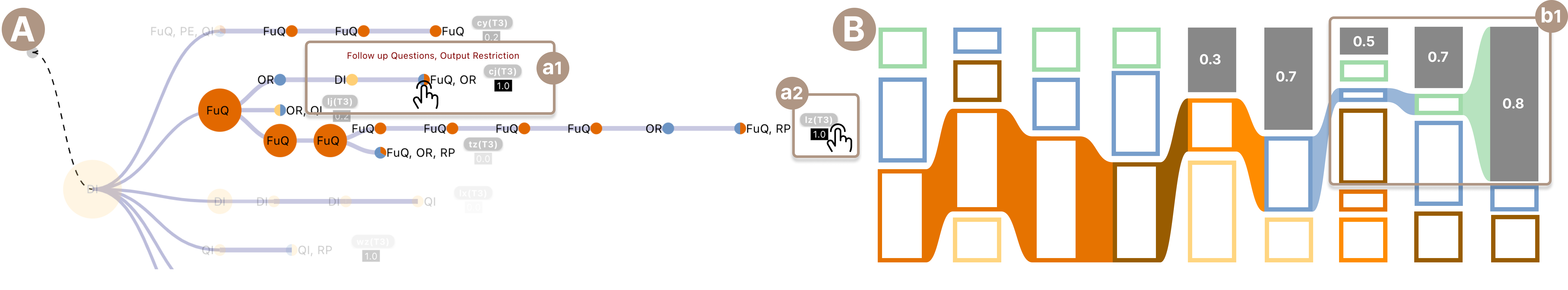}
    \caption{(A) The \textbf{Interaction Tree} visualization within \textit{Pattern Nuance} traces the detailed evolution of each student's interaction pattern for specific tasks. The students' paths utilizing $\{ 
 \text{``Follow up Questions''}\}$ pattern are highlighted. (B) An alternative design featuring a Sankey-stacked bar chart for tracking pattern evolution.}
    \label{fig:TreeDesign}
\end{figure*}

The \textit{Pattern Summary} is introduced with a control button at the top-left corner (\cref{fig:overallCards}) for selecting the grouping mode. A comprehensive macro-level pattern summary \textbf{(R2)} under the chosen mode is displayed below. The grey card component presents a between-group level summary (\cref{fig:overallCards}-A), adjacent to which are the within-group level summary cards (\cref{fig:overallCards}-B). To begin, instructors can view the summary donut chart (\cref{fig:overallCards}-a1), which shows the distribution of all students' prompt categories as defined in \cref{prompt_coding}. For example, here, a predominantly light yellow section suggests that a majority of student prompts are at the ``Remember'' cognitive stage, while minor blue and major green segments indicate a limited use of literature-supported effective prompt strategies. This donut chart design aims to clearly display the percentage of each cognitive level in students' prompts while optimizing space usage. Additionally, to effectively display and compare the overall quality of ChatGPT responses and students' learning outcomes, we selected two light grey bars and a dark grey bar (\cref{fig:overallCards}-a1) to represent the three metrics due to their simplicity and effectiveness \cite{munzner2014visualization}. The details of metrics are introduced in \cref{gpt_metrics}.\newline \indent Instructors can easily check each group's background information by hovering over the summary donut chart (\cref{fig:group comparison & hover}-A) and perform between-group comparisons of cognitive stage distribution and ChatGPT response quality using stacked bar charts and accompanying grey bar charts (\cref{fig:overallCards}-a2). Moreover, by switching to ``TaskG'' mode (\cref{fig:group comparison & hover}-B), instructors can change the grouping from student background to task types, shifting the analysis towards task-specific student performance and interaction summaries. This flexible approach enables an efficient multi-viewpoint summary of macro-level students-ChatGPT conversations (\textbf{R2}). After identifying a specific group of interest, instructors can click on a stacked bar in the summary card (\cref{fig:overallCards}-a2), which highlights the selected group card (\cref{fig:overallCards}-b1) for deeper, within-group analysis. Instructors can sort students by various metrics by clicking on the metric summary bars in the first row, as shown in the example where students are sorted by their scores in ascending order (\cref{fig:overallCards}-b1). To move on, instructors can click on any stacked bar or the donut chart (\cref{fig:overallCards}-b1) to investigate micro-level details in \textit{Pattern Mining Table} (\cref{fig:tableVis}-A) and the \textit{Interaction Tree} (\cref{fig:TreeDesign}-A).

\subsubsection{Pattern Mining Table}
The \textit{Pattern Mining Table} (\cref{fig:Teaser}-b4, \cref{fig:tableVis}-A) catalogs micro-level nuanced interaction patterns mined from the tasks and students selected by users (\textbf{R3}). Each interaction pattern is associated with specific metrics: length (``L''), frequency (``C''), and average score (``Avg.''). Aligned with the previous definition \cref{prompt_coding}, interaction patterns are delineated as either a set of codes (denoted by curly braces $\{\}$, \cref{fig:tableVis}-B) or an ordered list of codes (indicated by an arrow $\rightarrow$ in the ``Pattern'' column, \cref{fig:tableVis}-C), with each code's background color reflecting its category. Instructors can sort these patterns by any metric, aiding in the identification of prevalent or effective patterns.

\subsubsection{Interaction Tree}
The \textit{Interaction Tree} employs a decision-tree format design to trace the evolution of interaction patterns (\textbf{R4}) for individual students under each task (\cref{fig:Teaser}-b5, \cref{fig:TreeDesign}-A). Each path within the \textit{Interaction Tree} represents a student's detailed interaction process with ChatGPT under a specific task, beginning from a common root where nodes symbolize students' prompts, and adjacent solid links depict ChatGPT's responses. This format enables the aggregation of similar interaction paths, highlighting variations and commonalities in student interaction patterns. Consistent with the above settings, colors in the node coding for the category of the prompt and abbreviations inside or beside the node detailing the code content. When prompts encompass multiple codes, nodes incorporate pie-chart coloring to represent each code visually (\cref{fig:TreeDesign}-a1). To aggregate prompts with the same code contents, we use node size to indicate the number of students at the same round in their conversation. On the other hand, the solid links between nodes convey ChatGPT's response characteristics, including ``Response Token Length'' (RL) and ``Information Gain'' (IG). Each ChatGPT response's ``Information Gain'' is represented through variations in the horizontal length of the link, as it is the major information instructors (E1 \& E2) want to notice, while link width and opacity double encode ``Response Token Length''. This visual encoding offers insights into the value and quality of ChatGPT's replies. Furthermore, the end of each path features one grey tag representing the student's alias and task's ID (\cref{fig:TreeDesign}-a2), together with another tag below encoding the student's performance on this task. Here, we utilized numerical values and color intensity to denote scores, thereby linking interaction patterns directly to learning outcomes. For instance, the highlighted student gained the full mark (\cref{fig:TreeDesign}-a2) after various interactions with ChatGPT. Through these meticulously designed components, instructors are not only equipped to perform a granular analysis of student interactions but also able to correlate them with educational outcomes, facilitating a comprehensive understanding of students' learning behaviors and the effectiveness of their interactions with ChatGPT (\textbf{R5}).

To further facilitate in-depth evaluation of each identified interaction pattern, we introduced an interplay between the \textit{Pattern Mining Table} and the \textit{Interaction Tree}. By clicking on each row, which corresponds to a specific interaction pattern (\cref{fig:tableVis}-B), the corresponding students who engage in this pattern will be highlighted in the \textit{Interaction Tree} (\cref{fig:TreeDesign}-A). This enables instructors to assess relevant metrics such as learning outcomes and the quality of ChatGPT’s responses for each identified pattern or individual student (\textbf{R5}).

\textbf{Design alternative.} During the design process, some alternatives were raised and discussed with our experts. For instance, we proposed to use a Sankey diagram \cite{riehmann2005interactive} together with a stacked bar chart design to represent the students' different prompt choices at each step. Specifically, each stacked bar represent the distribution of different categories of codes in each conversation turn (\cref{fig:TreeDesign}-b1). The Sankey flow represents the selected student's interaction pattern, ending with a grey bar with numerical values representing the student's score in this task. Although this design was clear to show the percentage of students' diverse interaction choices with ChatGPT at each conversation turn, the experts (E1 \& E2) prioritized our Interaction Tree visualization since the quality of each ChatGPT's response was lacking and could be hard to add to the Sankey stacked bar chart easily. Meanwhile, they preferred a nuanced comparison between different students' interaction patterns, which is also a limitation of the Sankey-form diagram.

\subsection{Detail View}
Instructors can select the student's alias and task's ID tag at the end of each path in the \textit{Interaction Tree} (\cref{fig:TreeDesign}-a2) to examine the specifics of each task and students' responses in the \textit{Task Description} (\cref{fig:Teaser}-c1), and explore students' raw conversation with ChatGPT in the \textit{Raw Conversation} (\cref{fig:Teaser}-c2). This functionality not only validates the analytical findings but also equips instructors with concrete examples and references for crafting feedback to students or refining task designs, serving as the end of our whole analysis workflow (\textbf{R6}).

\section{Evaluation}
% In this section, we first show the students' questionnaire feedback as an evaluation of the rationality of our open-ChatGPT in-class exercise design. Then, we demonstrate the effectiveness of {\systemName} to help users identify and analyze students' interaction with ChatGPT through three case studies and interviews with six domain experts (\textbf{E1-E6}, one full professor, three assistant professors and two lecturers, from three different universities). We also summarized the feedback from all expert in this section. 

This section delves into the assessment of {\systemName}. We present the result of a student questionnaire to validate the settings of our open-ChatGPT in-class exercises. Then, we demonstrate the effectiveness of the Information Gain (IG) metric we introduced in \ref{gpt_metrics}. Subsequently, we showcase the system's capacity to facilitate the identification and analysis of student interactions with ChatGPT through three case studies. Additionally, we discuss the collective feedback from interviews with six domain experts (\textbf{E1-E6}).
% , encompassing one full professor, three assistant professors, and two lecturers from three universities. 
The insights from these experts evaluated the {\systemName}'s effectiveness and impact.

\subsection{Questionnaire Feedback}
% As we introduced in \textbf{Section Three}, the format and tasks we used in the in-class exercises section were defined and preliminary tested via a mode class section with 30 postgradue computer science students. To evaluate its rationality, we conducted a voluntary feedback questionnaire section in the middle of the semester. The questionnaire consists of three multiple choice questions and three short answer questions. The three multiple choice questions focus on students' experience and thoughts towards using ChatGPT to learning data visualization and do the in-class exercises we designed. The short answer questions collect students opinions towards their general comments, trust and feelings of using ChatGPT to learn data visualization and other subjects. The questionnaire results demonstrate that the majority of students enjoy the usage of ChatGPT in learning data visualization and in our in-class exercises and are willing to use it a lot. The questionnaire results proved the rationality of our course material design and guarantee the quality of data we collected. 

We administered a mid-semester voluntary questionnaire consisting of multiple-choice and short-answer questions. The multiple-choice questions were designed to gauge students' experiences and attitudes towards using ChatGPT to learn data visualization and complete the designed in-class exercises. The short-answer questions aimed to collect students' general feedback, including their level of trust in and feelings about using ChatGPT for learning data visualization and other subjects. The detailed statistics of the questionnaire are provided in the supplementary (B). To summarize, the results indicated a strong positive reception: more than $90\%$ students reported enjoying using ChatGPT in their learning process and expressed a willingness to utilize it extensively in our data visualization course. These findings affirmed the rationality behind our course material design and ensured the quality of the data collected for our study. Some other insights from the short-answer questions are discussed in the following \cref{sec:discussion}.

\subsection{Metric Evaluation}
\label{metric_evaluation}

We evaluated the effectiveness of our Information Gain (IG) metric by sampling 10\% of student-ChatGPT conversations. Two experts (E1 \& E2) manually labeled the data into three categories: ``low information gain'' (score 0), ``average information gain'' (score 0.5), and ``high information gain'' (score 1). A score of 0 was assigned to responses containing mostly incorrect information, a score of 1 to responses providing rich and accurate new information, and a score of 0.5 to responses that were partially inaccurate or partially redundant with previous. To measure the correlation between IG metric and experts' labeling, we calculated Pearson correlation \cite{sedgwick2012pearson}, Spearman correlation \cite{sedgwick2014spearman} and Kendall Rank correlation \cite{abdi2007kendall} between them. The results are 0.609 ($p=0.00$), 0.621 ($p=0.00$) and 0.497 ($p=0.00$), respectively.
All the results show there is a moderate to strong and significant positive correlation between the IG metric and experts' judgment of ChatGPT's response quality. The sample of labeled data and metric results is provided in the supplementary materials. Although effective, the IG metric is a simplified measure that primarily considers word frequency, designed to provide an initial assessment. In the future, we plan to use more advanced third-party ChatGPT response quality evaluation methods to enhance the metric accuracy.

\subsection{Case Study}
% We invited all our experts to explore the system independently. In the below sections, we present the three case studies summarized from experts' findings to demonstrate the effectiveness of our system.
We engaged experts (E1-E6) to evaluate {\systemName} independently. We introduced the background, visual designs, and a brief workflow demo to them and yielded three case studies that underscore the system's utility in analyzing student interactions with ChatGPT. Experts E1 and E2 are the course instructors we collaborated with, and E3-E6 are newly invited experts, including three assistant professors and one lecturer from three different universities, all with expertise in data visualization.

\subsubsection{Case 1: Enhancing Students' ChatGPT Utilization}
In the first case study, E1 and E3 leveraged {\systemName} to derive instructional strategies to optimize students' use of ChatGPT for learning and addressing challenging tasks.
% This case study focuses on formulating actionable instructions to aid students in leveraging ChatGPT for learning and problem-solving in challenging scenarios.

% \textbf{Initial Overview and Task Filtering:} The investigation commenced with an overview of tasks and student backgrounds. To nevigate the challenging tasks, the experts utilized the \textit{Task Overview}, to filter out tasks with difficulty scores below 3, together with those categorized under ``self-learning'' and ``remember'' and those with average scores exceeding 0.8 (\cref{fig:Teaser}(\textbf{a1})). In the \textit{Student Overview}, all students were retained to ensure a comprehensive analysis across diverse interaction patterns (\cref{fig:Teaser}(\textbf{a2})).

\textbf{Initial Overview and Task Filtering:} The investigation began with an overview of tasks and student backgrounds. To focus on challenging tasks, the experts used the \textit{Task Overview} to filter out tasks with difficulty scores below 3, those categorized under "self-learning" and "remember," and those with average scores exceeding 0.8 (\cref{fig:Teaser}(\textbf{a1})). In the \textit{Student Overview}, all students were retained to ensure a comprehensive analysis of diverse interaction patterns (\cref{fig:Teaser}(\textbf{a2})).

% Experts commenced by reviewing task and student backgrounds, utilizing the Task Overview to isolate tasks deemed challenging—those with difficulty scores below 3, categorized under "self-learning" and "remember", and possessing average scores above 0.8. This filtering aimed to pinpoint tasks where optimal student interaction strategies might emerge. All students were retained in the analysis to encompass a broad spectrum of interaction behaviors.

\textbf{Identifying Challenges and Specific Tasks:} To review task summary patterns, the experts switched to the ``Task-Grouping'' mode in the \textit{Pattern View} (\cref{fig:Teaser} \textbf{b1}) and identified ``analyze'' tasks as particularly challenging, evidenced by longer stacked bars indicating higher cognitive engagement (\cref{fig:Teaser} \textbf{b2}). Although the thick grey bar representing ``Information Gain'' suggested students acquired significant information from ChatGPT, the overall learning outcomes for these tasks were suboptimal (\cref{fig:Teaser} \textbf{b2}). Consequently, the experts focused on the ``analyze'' task group, identifying ``Task 3'' as notably difficult due to the extensive cognitive processing it required(\cref{fig:Teaser} \textbf{b3}).

\begin{figure}[!htb]
    \centering
    \includegraphics[width=\linewidth]{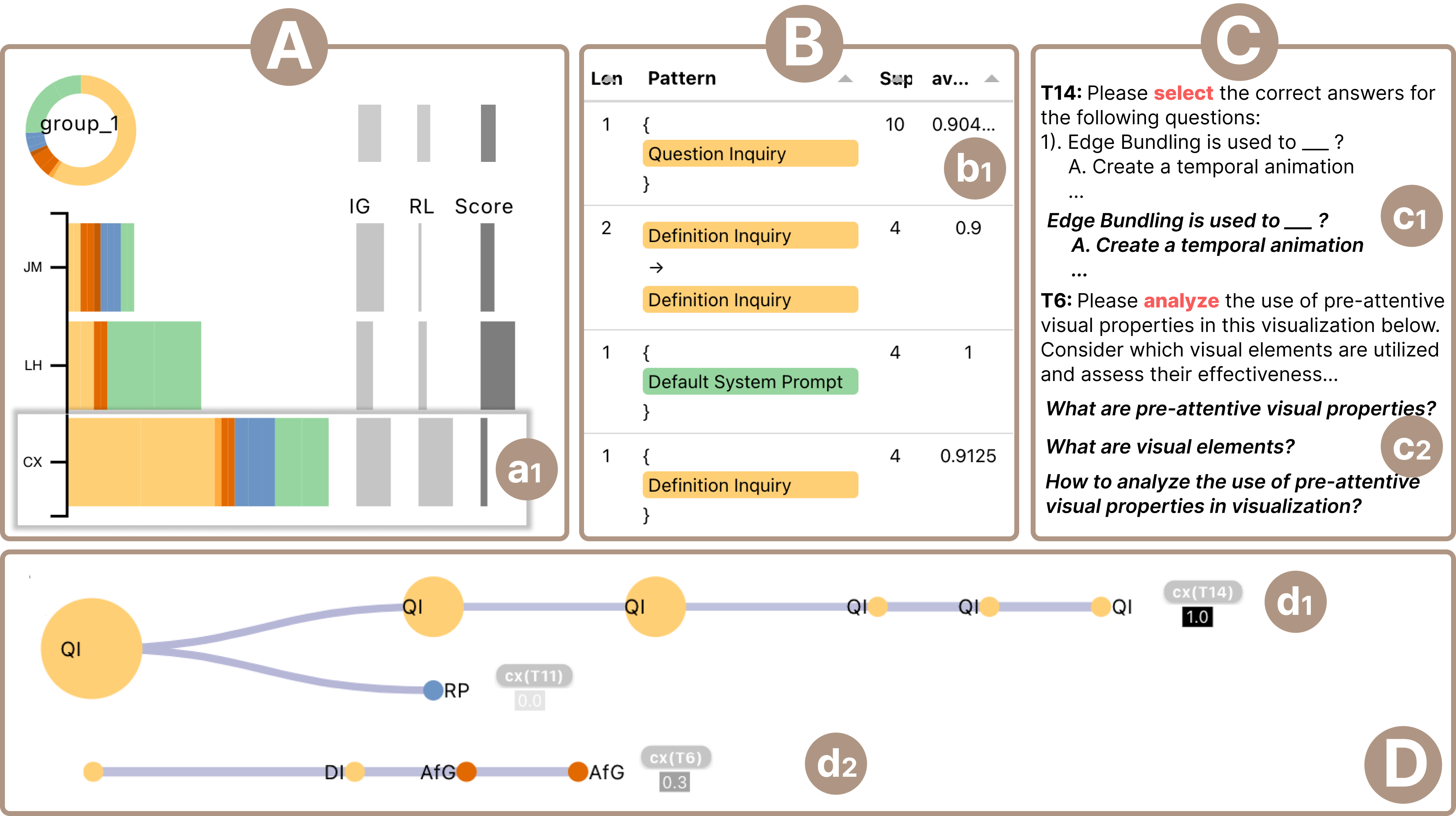}
    \caption{ (A) The interaction pattern summary within Group 1. (B) The pattern table, sorting all interaction patterns mined from students with alias ``cx''. (C) The descriptions for T14 \& T6 and the student's key prompts. (D) The pattern paths of student ``cx'' for T14 and T6.}
    \label{fig:case 2}
\end{figure}

\textbf{Analysis of Interaction Patterns:} The examination of ``Task 3'' through the pattern mining table and Interaction Tree (\cref{fig:Teaser}-b4, b5) revealed nuanced dynamics of student-ChatGPT interactions. Initially, sorting by average score revealed infrequent and overly specific patterns. To uncover broadly applicable patterns, experts shifted to sorting by frequency (``Count''), identifying a recurring and effective pattern: a combination of ``Definition Inquiry'' and ``Follow up Questions'' (\cref{fig:Teaser}-b6), notable for its prevalence and high average score exceeding 0.8. Experts agreed this simple yet effective combination was crucial for foundational understanding and deeper exploration of ChatGPT's responses. To observe students using this pattern, experts selected it in the table, highlighting corresponding students in the Interaction Tree (\cref{fig:Teaser}-b8, \cref{fig:TreeDesign}-A). A standout instance involved a student, ``CJ'', who used this pattern to achieve a full score with minimal conversation (\cref{fig:Teaser}-b8). By clicking the student-task ID tag (\cref{fig:Teaser}-b8) and reviewing the raw conversations (\cref{fig:Teaser}-c2) along the tree branch, experts saw ``CJ'' begin with a ``Definition Inquiry'', asking if ChatGPT understood the pipeline concept \cite{6875967}. After confirmation, ``CJ'' asked a ``Follow up Question'' for further exploration. Noticing that ChatGPT did not provide answers from the multiple choices listed in the task description, ``Output Restrictions'' were strategically implemented. Despite these restrictions, ChatGPT's responses remained flawed, leading ``CJ'' to revert to ``Definition Inquiry'' for double verification before progressing. Once a correct response was secured from ChatGPT, ``CJ'' advanced through the remaining sections, receiving partially correct responses from ChatGPT but ultimately providing completely correct answers independently. Experts concluded that ``Output Restrictions'' in prompts maximize ChatGPT's utility for precise information retrieval. Revisiting the pattern mining table (\cref{fig:Teaser}-b4), they noted ``Output Restrictions'' as a significant pattern (\cref{fig:Teaser}-b7) due to its high frequency and average score. Thus, the experts finally identified the combination of ``Definition Inquiry'' and ``Follow up Questions'' with ``Output Restrictions'' as a potent strategy for engaging with ChatGPT effectively. The raw conversation data of ``CJ'' is provided in the supplementary (C). This evaluation also highlighted ChatGPT's limitations in addressing abstract questions, advising against outright reliance on its answers but encouraging a focus on its reasoning and factual accuracy. Ultimately, the experts planned to incorporate these insights into feedback for their students.

% The experts concurred that incorporating ``Output Restrictions'' into prompts is an effective strategy for students to maximize the utility of ChatGPT in accessing precise information. Revisiting the pattern mining table (\cref{fig:Teaser}-b4), they noted that ``Output Restrictions'' also stood out as a significant pattern (\cref{fig:Teaser}-b7), characterized by high frequency and average score, validating it as a commendable approach that guides ChatGPT to offer targeted and relevant responses. Through a meticulous analysis of these interactions, the experts identified the combination of ``Definition Inquiry'' and ``Follow up Questions'' with ``Output Restrictions'' as a potent strategy for engaging with ChatGPT effectively. Additionally, the evaluation highlighted ChatGPT's limitations in addressing abstract questions, advising against outright reliance on its answers but rather encouraging a focus on its reasoning and the factual accuracy of its responses. Ultimately, the experts planned to incorporate these insights into feedback for their students.

This case study showcases {\systemName}'s capability to not only unearth effective student interaction patterns with ChatGPT but also distill these insights into actionable strategies for educators.

\subsubsection{Case 2: Provide Personalized Feedback}

Experts E3 and E4 aimed to provide personalized feedback to students lagging behind, focusing on those with average scores below 0.5 in the \textit{Student Overview}.
They identified ``Group 1'' from the \textit{Pattern View} donuts chart as the weakest, with student ``cx'' showing extensive ChatGPT interaction but the lowest scores (\cref{fig:case 2}-a1). 
% It demonstrated that ``cx'' extensively interacted with ChatGPT but unfortunately scored the lowest within the group (\cref{fig:case 2}-A).

% Experts E3 and E4 sought to extend personalized learning feedback to students who were lagging behind their peers. Their approach entailed retaining all task types within the system while filtering out students in the \textit{Student Overview} who had an average score above 0.5. 
% This strategy aimed to focus on students facing the most significant challenges. From the donuts chart in the \textit{Pattern View}, ``Group 1'' was identified as having the weakest background and performance. 
% After a closer examination of this group, experts identified a student, anonymously named ``cx'', who had a notably long usage bar but the lowest ``Score'' indicator (\cref{fig:case 2}-a1). It demonstrated that ``cx'' extensively interacted with ChatGPT but unfortunately scored the lowest within the group (\cref{fig:case 2}-A).

To better understand ``cx's'' interactions, the experts analyzed the \textit{Pattern Table} (\cref{fig:case 2}-B) sorted by usage frequency (``Count''), revealing ``cx'' predominantly engaged in basic ``Question Inquiry'' and ``Definition Inquiry'' and rarely modified default settings (\cref{fig:case 2}-b1). This pattern indicated that ``cx'' primarily operated at initial cognitive levels, showing a dependency on ChatGPT's capabilities rather than engaging in self-driven learning or critical thinking. 
% However, alongside the system's workflow, experts' further analysis illuminated a distinct contrast in the student identified as ``cx's'' engagement with various tasks. 

% To understand ``cx's'' interaction dynamics with ChatGPT, experts sorted the \textit{Pattern Table} (\cref{fig:case 2}-B) by frequency of use (``Count''). The results showed that ``cx'' predominantly used basic ``Question Inquiry'' and ``Definition Inquiry'', meanwhile seldom modify the default ChatGPT settings (\cref{fig:case 2}-b1). The experts thus concluded that this student mainly stayed at initial cognitive levels and also lacked diverse prompt strategies. The patterns shown suggested a reliance on ChatGPT's capabilities rather than engaging in self-driven learning or critical thinking. However, alongside the system's workflow, experts' further analysis illuminated a distinct contrast in the student identified as ``cx's'' engagement with various tasks. 

However, further analysis showed a distinct contrast in ``cx's'' engagement with various tasks. In task T14, consisting of straightforward multiple-choice questions, ``cx'' achieved full marks by directly copying questions into ChatGPT (\cref{fig:case 2}-c1, d1).
% In Task 14 (T14), composed of straightforward multiple-choice questions, the experts saw ``cx'' effortlessly achieving full marks by posing direct questions (question copy \& paste) to ChatGPT (\cref{fig:case 2}-c1, d1). 
While effective for securing marks, this approach had limited educational value as it bypassed the learning process. In contrast, despite the lower score in T6, ``cx'' demonstrated a significant effort to grasp the underlying concepts and principles of analysis, as evidenced by the detailed prompts samples (\cref{fig:case 2}-c2) and the orange nodes in the interaction path (\cref{fig:case 2}-d1). 
% However, this approach, while effective for securing marks, was not favored by experts due to its limited educational value, as it bypasses the learning process entirely. 
% In stark contrast, despite the lower score in T6, ``cx'' demonstrated a significant effort to grasp the underlying concepts and principles of analysis, as evidenced by the detailed prompts samples in (\cref{fig:case 2}-c2) and the orange nodes in the interaction path in (\cref{fig:case 2}-d1). 

The experts identified that T6 was an exercise in ``visualization analysis and evaluation'' and speculated that ``cx'' had a keen interest in tasks requiring analysis and evaluation rather than just recalling basic concepts. Consequently, the experts proposed a shift in ``cx's'' educational strategy, advocating for an increased focus on tasks that emphasize analysis and evaluation and moving away from tasks that merely require regurgitating information.

% Meanwhile, the experts deduced that the student's struggle with T6 stemmed from a foundational gap in knowledge rather than a lack of initiative or understanding of the task's analytical demands.
% Consequently, the experts proposed a shift in ``cx's'' educational strategy, advocating for an increased focus on tasks that emphasize analysis and evaluation and moving away from tasks that merely require regurgitating information. 

This case demonstrates the power of {\systemName} to aid instructors in providing in-depth personalized feedback to students. 

\begin{figure}[!htb]
    \centering
    \includegraphics[width=\linewidth]{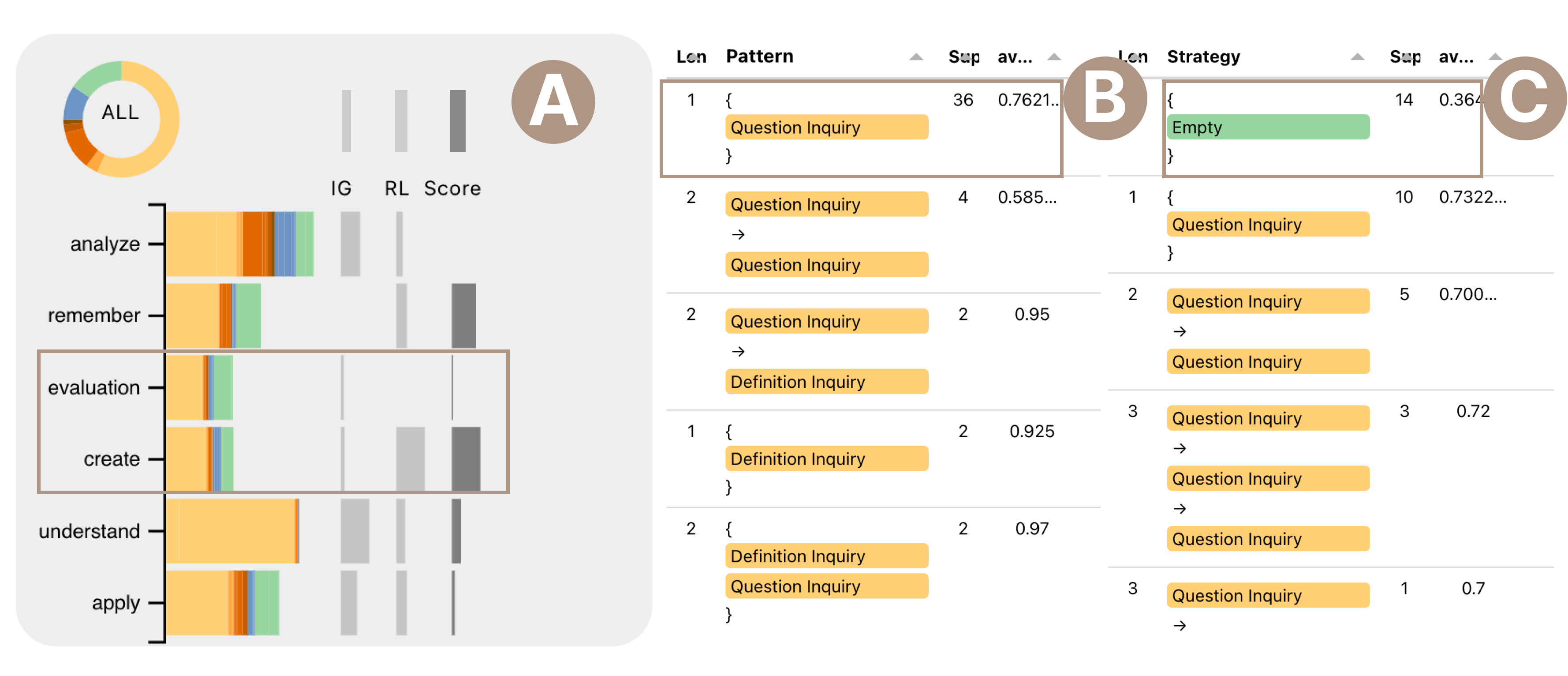}
    \caption{ (A) Summary of students' interaction patterns for each type of task. (B) The pattern table, sorting all interaction patterns mined from students under the ``evaluate'' type task. (C) The pattern table, sorting all interaction patterns mined from students under the ``create'' type task.}
    \label{fig:case 3}
\end{figure}    

\subsubsection{Case 3: Refine Course Material Design}

Experts E2 and E4 employed {\systemName} to assess the alignment of existing in-class exercises with educational objectives, especially considering unrestricted access to ChatGPT. Their initial step involved excluding ``self-learning'' tasks from the analysis within the \textit{Task Overview}.

Progressing to the \textit{Pattern View}, they aimed to compare cognitive engagement levels across various task types. The experts found that ``evaluate'' and ``create'' tasks elicited notably lower levels of advanced cognitive engagement compared to ``analyze'' and ``remember'' tasks (\cref{fig:case 3}-A). They further examined the ``evaluate'' tasks and identified a predominant pattern of ``Question Inquiry'' (\cref{fig:case 3}-B), indicating students often relayed task descriptions to ChatGPT for answers. A detailed check of the ``evaluate'' task descriptions in the \textit{Detailed View} revealed this was particularly evident in tasks where students were explicitly instructed to analyze specific visual properties. The direct provision of analysis criteria encouraged a straightforward query-response dynamic with ChatGPT, bypassing deeper cognitive processing.

% After a detailed check of task descriptions of each ``evaluate'' type task in the \textit{Detailed View}, the experts identified that this was particularly evident in tasks where students were explicitly instructed to analyze specific visual properties. 

% The direct provision of analysis criteria seemingly encouraged a straightforward query-response dynamic with ChatGPT, bypassing deeper cognitive processing. 
Based on these insights, E4 recommended a shift in exercise design towards more open-ended questions that prompt students to independently determine relevant metrics or principles before engaging in analysis. This approach aims to ensure alignment with educational objectives considering unrestricted access to ChatGPT. On the other hand, the experts investigated ``create'' type tasks and found that the analysis results and task description indicated a lack of clear direction for students in formulating queries to ChatGPT, often resulting in incomplete (``Empty'') or superficial (``Question Inquiry'', mainly refer to question copy \& paste) interactions (\cref{fig:case 3}-C). To address this, it was suggested that instructors could guide students toward requesting alternative design options from ChatGPT, including the rationale behind each. The experts concurred that prompting students to assess these options and incorporate their critical reasoning can greatly enhance the learning experience.

% This case proved the effectiveness of {\systemName} for aiding instructors do course material redesigns under the new ``variable'' ChatGPT involved to ensure the educational objectives could be achieved. 

This case illustrates the effectiveness of {\systemName} in supporting instructors with the redesign of course materials, particularly in integrating ChatGPT, to guarantee the achievement of educational objectives.

\subsection{Expert Interview}

Following the case studies, one-on-one interviews were conducted with six experts, each lasting around 80 minutes with a \$80 compensation.

\textbf{System Workflow.} The workflow of {\systemName} was praised by all experts for its clarity and functionality. Experts noted its ease of use, allowing for a streamlined narrowing of analysis scope to achieve insights across multiple levels. Whether focusing on groups of students, individuals, or tasks, the system's design facilitated seamless navigation without added complexity. E5, an assistant professor specializing in teaching visualization to business students, highlighted, \textit{``The workflow's logical progression and the interconnection of each view were particularly impressive, enabling a diverse analytical focus through a unified procedure.''} E3, E4, and E6 emphasized the importance of understanding students' learning outcomes. They appreciated the system's capability to filter, and highlight scores at various analytical stages.

\textbf{Visual Design and Interactions.} Experts agreed that the visual design and interactive elements of {\systemName} are clear and user-friendly, significantly enhancing the analytical process. The use of stacked bar charts was particularly commended for facilitating easy understanding and comparison of cognitive levels across students. The color coding, distinguishing between learning-related and ChatGPT usage-related codes, was found intuitive. E2 highlighted the clarity provided by the visual design: \textit{``The ability to discern students' overall cognitive level at a glance is highly appreciated.''} The Interaction Tree visualization emerged as a favorite for its detailed representation of different patterns and the entire interaction journey, including ChatGPT responses and learning outcomes. E4, an assistant professor, praised the decision-tree format for showcasing diverse student strategies and ChatGPT's varied response quality. However, concerns were raised about the scalability of this visualization, especially for large classes over 100 students, suggesting a need for refining the summary of popular interaction patterns and detailed comparison of individual paths.

\textbf{Suggestions.} Experts provided several actionable suggestions for enhancing {\systemName}. E5 proposed adding a summary report panel to capture screenshots and annotate findings directly within the system, facilitating a comprehensive and customizable analysis experience. E6 recommended more flexible options for grouping tasks and students, suggesting a user-defined grouping mechanism to enable richer cross-correlations between different cohorts and tasks, visualized through a matrix-form panel. Despite the potential for increased complexity, E6 believed this feature could unveil deeper insights. Additionally, E1 and E4 suggested integrating the coding of students' conversations with ChatGPT directly into the workflow, similar to a ``grading the assignment'' process. This would streamline the evaluation process and enrich instructors' understanding of student interactions with ChatGPT.

\section{Discussion}
\label{sec:discussion}
This section discusses the significance and insights towards ChatGPT for data visualization education, and the generalizability and scalability of the proposed visual analytics system.

\textbf{ChatGPT for Data Visualization Education:} The introduction of ChatGPT into educational ecosystems marks a pivotal moment and necessitates a nuanced understanding of technological integration in learning. Our investigation into the patterns and strategies of student engagement with ChatGPT is critical, providing insights that guide instructors in facilitating the effective use of AI. Our study revealed several key findings from student questionnaires, student-ChatGPT conversation data, and discussions with course instructors. First, students reported that ChatGPT excels in summarizing key concepts, providing quick access to vast information, and offering tailored Q\&A sessions. Instructors agreed that these functionalities are particularly beneficial for students with limited backgrounds, such as those who changed their major in graduate school, allowing them to keep pace with coursework without hindering class progress. Additionally, over $90\%$ of students expressed satisfaction with ChatGPT's ability to handle data visualization queries, indicating a strong positive perception of its utility in data visualization education. However, instructors pointed out that recognizing ChatGPT's limitations in interpreting and processing visual data compared to textual information is crucial. For instance, through the collected student-ChatGPT conversations, instructors identified that some students resorted to asking ChatGPT for URLs of existing sample visualizations (e.g., demos of parallel coordinates) to obtain better figure quality after receiving a low-quality response. These findings inspire instructors to develop innovative approaches, such as guiding ChatGPT to provide descriptions or links to reference visualizations, for better educational content. Instructors also emphasized that the demand-driven nature of data visualization requires students to employ precise and strategic questioning techniques. This underscores the importance of teaching students how to prompt effectively to maximize ChatGPT's capabilities. Furthermore, as we navigate the ChatGPT-enhanced educational paradigm, our work highlighted the need to focus on cultivating students' higher-order cognitive skills, such as critical thinking and evaluative judgment. For instance, instructors suggested encouraging students to ask ChatGPT for alternative designs with underlying rationales, enabling them to critically assess options and fostering a collaborative learning dynamic. Approaches like this redefine the role of ChatGPT in education—from a universal solver to a pedagogical partner—and emphasize the importance of critical engagement and decision-making skills in the data visualization domain.

\textbf{Generalizability \& Scalability:} The application of {\systemName}, while rooted in the context of data visualization education, unveils broader implications for ChatGPT-assisted learning environments. It offers a robust framework for analyzing student interactions with ChatGPT across a variety of courses. {\systemName}'s core workflow, which includes filtering tasks and students of interest, analyzing cognitive levels through prompts, assessing prompt engineering skills, evaluating ChatGPT’s response quality, and identifying interaction patterns, encapsulates the universal aspects of leveraging ChatGPT in education. This approach enriches our understanding of effective AI integration into pedagogy and opens avenues for examining ethical considerations regarding ChatGPT's involvement in teaching practices. 
Regarding scalability, {\systemName} demonstrates competence in managing classes of 48 students through the interaction tree visualization techniques equipped with simple pruning for clarity. This capability ensures the {\systemName}'s efficacy in distilling actionable insights from complex datasets, catering to the needs of regular-sized classes. However, scalability challenges arise as class sizes expand beyond this scope. For larger cohorts (e.g., exceeding 100 students) the incorporation of edge bundling techniques emerges as a potential refinement to enhance pattern visualization and comparison. Additionally, expanding the pattern mining table to include interaction patterns from different tasks or courses would further enhance its scalability and provide additional benefits. This adaptation will form a component of our further efforts, aiming to ensure that {\systemName} remains effective in in diverse educational settings, advancing the goal of inclusive AI-enhanced education.

\section{Conclusion}
This study introduces {\systemName}, a visual analytics system for instructors to analyze student-ChatGPT interactions. In particular, we collected student-ChatGPT conversations in a graduate-level data visualization course and developed a comprehensive coding scheme to categorize students' prompts from cognitive levels and ChatGPT's response qualities. We then build {\systemName} to visualize student-ChatGPT interaction patterns and support multi-level, multi-perspective analysis. {\systemName} empowers instructors with deep insights into students' cognitive processes, their reliance on ChatGPT, and their ability to use it effectively, highlighting areas for pedagogical intervention to promote higher-order thinking. The system's effectiveness, validated through expert interviews and case studies, confirms its potential to impact student-ChatGPT conversation analysis and visualization education. As we look to the future, {\systemName} sets the stage for broader research into the application of visual analytics in education and the development of AI-enhanced personalized learning experiences. 

\acknowledgments{
This work has been partially supported by ITF grant PRP/017/22FX and RGC GRF grant 16218724. 
}

\bibliographystyle{abbrv-doi-hyperref}

\bibliography{template}

\appendix % You can use the `hideappendix` class option to skip everything after \appendix
% \appendix % You can use the `hideappendix` class option to skip everything after \appendix

% \section{About Appendices}
% Refer to \cref{sec:appendices_inst} for instructions regarding appendices.

% \section{Troubleshooting}
% \label{appendix:troubleshooting}

% \subsection{ifpdf error}

% If you receive compilation errors along the lines of \texttt{Package ifpdf Error: Name clash, \textbackslash ifpdf is already defined} then please add a new line \verb|\let\ifpdf\relax| right after the \verb|\documentclass[journal]{vgtc}| call.
% Note that your error is due to packages you use that define \verb|\ifpdf| which is obsolete (the result is that \verb|\ifpdf| is defined twice); these packages should be changed to use \verb|ifpdf| package instead.

% \subsection{\texttt{pdfendlink} error}

% Occasionally (for some \LaTeX\ distributions) this hyper-linked bib\TeX\ style may lead to \textbf{compilation errors} (\texttt{pdfendlink ended up in different nesting level ...}) if a reference entry is broken across two pages (due to a bug in \verb|hyperref|).
% In this case, make sure you have the latest version of the \verb|hyperref| package (i.e.\ update your \LaTeX\ installation/packages) or, alternatively, revert back to \verb|\bibliographystyle{abbrv-doi}| (at the expense of removing hyperlinks from the bibliography) and try \verb|\bibliographystyle{abbrv-doi-hyperref}| again after some more editing.

\end{document}